\documentclass[12p,twocolumn,prl,a4paper,showpacs,superscriptaddress,amsmath,amssymb]{revtex4}

\usepackage{graphicx}
\usepackage{dcolumn}

\newcommand{\compA}{$\mathrm{Ga}_{1-x}\mathrm{Mn}_{x}\mathrm{As}$}
\newcommand{\compB}{(Ga,Mn)As}
\newcommand{\compC}{\mbox{\textit{a}-As}}
\newcommand{\mnga}{$\mathrm{Mn}_{Ga}$}
\newcommand{\mni}{$\mathrm{Mn}_{I}$}
\newcommand{\asga}{$\mathrm{As}_{Ga}$}
\newcommand{\tc}{$T_{C}$}
\newcommand{\Ta}{$T_{a}$}
\newcommand{\ta}{$t_{a}$}
\newcommand{\tg}{$T_{g}$}
\begin{document}
%
    \title{Influence of annealing parameters on the ferromagnetic properties of optimally passivated \compB\ epilayers}

    \author{V. Stanciu}
        \email{victor.stanciu@angstrom.uu.se}
        \affiliation{Deptartment of Engineering Sciences, Uppsala University, Box 534, SE-751 21 Uppsala, Sweden}
    \author{O. Wilhelmsson}
            \affiliation{Deptartment of Material Chemistry, Uppsala University, Box 538, SE-751 21 Uppsala, Sweden}
    \author{U. Bexell}
            \affiliation{Dalarna University, SE-781 88 Borl\"ange, Sweden}
    \author{M. Adell}
            \affiliation{Department of Experimental Physics, Chalmers University of Technology, SE-41296 G\"oteborg, Sweden}
    \author{J. Sadowski}
            \affiliation{Institute of Physics, Polish Academy of Sciences, PL-02668 Warszawa, Poland}
    \author{J. Kanski}
            \affiliation{Department of Experimental Physics, Chalmers University of Technology, SE-41296 G\"oteborg, Sweden}
    \author{P. Warnicke}
            \affiliation{Deptartment of Engineering Sciences, Uppsala University, Box 534, SE-751 21 Uppsala, Sweden}
    \author{P. Svedlindh}
        \affiliation{Deptartment of Engineering Sciences, Uppsala University, Box 534, SE-751 21 Uppsala, Sweden}


\date{\today}
    \begin{abstract}
The influence of annealing parameters - temperature (\Ta) and time
(\ta) - on the magnetic properties of As-capped \compB\ epitaxial
thin films have been investigated. The dependence of the transition
temperature (\tc) on \ta\ marks out two regions. The \tc\ peak
behavior, characteristic of the first region, is more pronounced
for thick samples, while for the second (`saturated') region the
effect of \ta\ is more pronounced for thin samples. A right choice
of the passivation medium, growth conditions along with optimal
annealing parameters routinely yield \tc-values of $\sim 150$ K
and above, regardless of the thickness of the epilayers.
    \end{abstract}
    \pacs{75.50.Pp, 75.30.Gw, 75.70.-i}
    \maketitle
%
Tremendous experimental and theoretical work has recently been
dedicated to the study of III-V diluted magnetic semiconductors
(DMS). Particularly \compB\ has received sustained interest
because of its relatively high ferromagnetic transition
temperature, being regarded as the prototype of this new class of
materials \cite{Ohno:SolidStComRev,Samarth:NM-RevGaMnAs2005}.
However, device applications call for further improvements of its
magnetic properties.

In this compound, the Mn impurities ideally substitute on Ga sites
and occupy random positions in the GaAs matrix. The substitutional
Mn cation (\mnga) has a double role in the lattice host; it acts
as a magnetic ion (its five $3d$ electrons do not participate in
the bonding) and as an acceptor yielding one hole in the GaAs
matrix. The magnetic interactions between the magnetic ions are
mediated by the free holes. A common point in all theoretical
descriptions of ferromagnetism in these systems is that the
ferromagnetic (FM) order has a direct correspondence to the net
dopant and hole concentration, $x$ and $p$, respectively. The
latter is very sensitive to the level of compensating defects.
High Mn concentrations require the MBE growth to be performed at
low temperatures (LT) and, due to this special growth condition,
native LT-GaAs defects \cite{Liu:APL-NativeDefLTGaAs1995} are
incorporated during the deposition process. One such defect is the
As antisite (\asga), i.e., As atoms substituting in the Ga
sublattice. \asga\ is a double donor. Hence, it compensates two
\mnga\ and therefore has a negative impact on the ferromagnetic
properties \cite{Korzhavyi:PRL-AsgaDef2002}. Mn atoms sitting in
the tetrahedrally As-coordinated interstitial positions (\mni)
\cite{Glas:PRL-DetMnIAndAsga2004} in the zinc-blende structure of
GaAs form another important type of defect
\cite{Maca:PRB-MnI-MnGa2002,Yu:PRB-LocOfMni2002}. \mni\ is also a
double donor \cite{Pethukov:PRL-SelfComp2002} and besides this,
its $d$ orbitals do not hybridize with the $p$ states of the holes
in the valence band and therefore do not participate in the
hole-induced ferromagnetism \cite{Blinowski:PRB-SpinMnI2003}.
Moreover, it was shown in Ref.~\cite{Blinowski:PRB-SpinMnI2003}
that \mni\ couples antiferromagnetically by a superexchange
mechanism with the nearest \mnga. Both \asga\ and \mni\ cause a
decrease of the hole concentration and the manganese magnetic
moment \cite{Korzhavyi:PRL-AsgaDef2002,Wang:JAP-95-6512-2004}.

A key issue for the improved magnetic properties recently reported
\cite{Edmonds:HighTcByResMonitAnn,Chiba:EffLTAnnTri2003,
Ku:APL-HighTcGaMnAs2003,Edmonds:MnIDiffusion} is the
\textit{`optimal'} LT annealing. There are three important
annealing parameters; the annealing environment, the annealing
time (\ta) and the annealing temperature (\Ta). In most of the
annealing studies performed hitherto, only one of these parameters
is varied, while the others are considered \textit{`optimal'}.
Because of these narrow experimental parameter windows there is no
consensus on what should be regarded as \textit{optimal} annealing
conditions. Concerning the annealing environment, annealings have
been performed either in-situ (in the deposition chamber,
immediately after the growth) \cite{Pross:JAP-InfLTAnnMicro2004}
or under the flow of a gas stream, e.g. pure N$_{2}$
\cite{Hayashi:APL-LTAnnGaMnAs2001,Potashink:APL-AnnEffGaMnAs2001}
or O$_{2}$ \cite{WVanRoy:cm-AnnInO2004}, or in air
\cite{Chiba:EffLTAnnTri2003,Edmonds:MnIDiffusion}. Most of these
procedures, particularly if performed at \textit{low} \Ta, are
lengthy, i.e., they require extended annealing times to yield
noticeable improvements of the magnetic properties. The annealing
temperatures are regarded as \textit{`low'} (e.g., 170 $^\circ$C
\cite{Edmonds:HighTcByResMonitAnn}) or \textit{`high'} (e.g., more
than 300 $^\circ$C \cite{Hayashi:APL-LTAnnGaMnAs2001}) when
compared to the growth temperature (\tg). Therefore, when defining
a reference for \Ta, one should always take into consideration the
growth conditions, especially \tg\ \cite{Victor:ExplTg}. The
annealing time, somehow following the conditions `\textit{low} \Ta
-- \textit{long} \ta' or vice versa, has been varied between a few
minutes to many days. Here, we show that a proper balance between
\Ta, \ta\ and, not the least, the annealing environment may give a
sense to what optimal annealing conditions are.

The mechanism behind the large increase of \tc\ (of the as-grown
samples) is the out-diffusion of the highly mobile \mni\
\cite{Yu:PRB-LocOfMni2002} to the free surface of the \compB\
epilayer, followed by their passivation in this region
\cite{Edmonds:MnIDiffusion}. Here, we choose as a reactive medium
a comparably thick cap layer of amorphous As (\compC). Passivation
of the freshly grown surfaces by capping with an amorphous As
layer is a well known procedure for pure semiconductors
\cite{Kowalczyk:JVST-AsPass1981,Heinlein:JVST-AsCap1999}. In case
of \compB, the efficiency of the \compC\ cap as regards \mni\
passivation has recently been addressed in Ref.~\cite{Adell:APL}.

This study focuses on \compA\ layers with Mn concentrations of
$x\sim0.05$ and $0.06$. The chosen range of $x$ has proved to give
the best magnetic properties \cite{Edmonds:HighTcByResMonitAnn}.
The study does not cover the ultrathin thickness range, nor does
it cover very thick layers - the film thickness in this study
varies between 100 \AA~to~1000 \AA. LT-MBE growth was employed and
for the range of doping concentrations mentioned above, \tg\ was
about 230 $^\circ$C. Further details on the sample preparation are
found elsewhere \cite{Adell:APL}. The $1 \times 1$ cm$^{2}$
as-grown samples were cleaved in many small pieces necessary for
subsequent post-growth annealings. The LT annealings were
performed in air in a closed oven (accuracy $\pm 3\ ^\circ$C). The
annealing temperatures (either higher or lower than \tg) ranged
from 150 $^\circ$C to 300 $^\circ$C.

Time-of-flight secondary ion mass spectrometry (ToF-SIMS) was
performed with a PHI TRIFT II instrument using a pulsed $15$ kV liquid
metal ion source enriched in $^{69}$Ga$^{+}$ ions. Depth profiles were
obtained by sputtering a surface area of $50 \times 50 \mu$m$^2$ and
measure on an analysing area of $25 \times 25 \mu$m$^2$. To enhance the
secondary ion emission the depth profiles were measured under
oxygen flooding with a base pressure of $10^{-6}$ Torr in the sample chamber.
The sputter-rate was determined from the full-width at half-maximum of the
Mn-profile for the \compB~layer.

Chemical analyses of the films were performed by X-ray photoelectron
spectroscopy (XPS) using a Phi Quantum 2000 instrument with monochromatizied
Al K$\alpha$ radiation with an accuracy of $\pm 0.1$ eV. Peaks
from As($3d$), Ga($2p$), Mn($2p$) and Mn($3p$) were analysed to settle the
chemical state. To compensate any drift in the binding energy between the
samples, due to charging effects, the C($1s$) peak was measured for all
films and positioned at $285.1$ eV.

\begin{figure}
        \centering
        \includegraphics[width=0.48\textwidth]{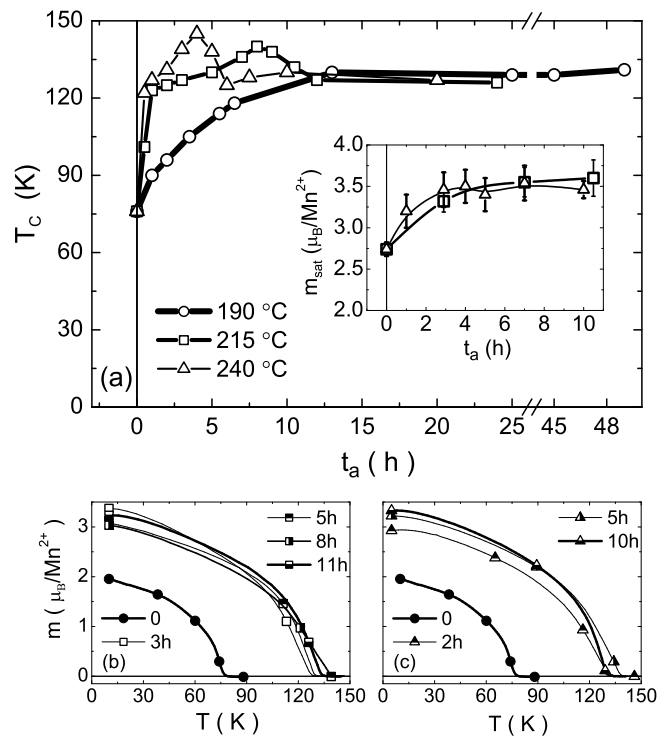}
        \caption{\label{fig:fig1}(a) \tc\ vs.
         annealing time for the 1000 \AA~thick film. The inset shows the variation
         of the saturation magnetic moment with \ta. (b) and
         (c) show the Brillouin-like $M(T)$ curves
         recorded at $T_{a}=215 ^\circ$C and $240 ^\circ$C,
         respectively (the as-grown sample is denoted by `0').}
\end{figure}

The magnetic measurements were performed in a QuantumDesign
MPMS-XL Superconducting QUantum Interference Device (SQUID)
magnetometer. The ferromagnetic transition temperatures were
derived from the temperature dependence of the magnetization,
$M(T)$, as the onset of the FM order. For consistency, the $M(T)$
measurements were performed in a low magnetic field (20 Oe)
(field-cooled protocol) applied in the film-plane along either
[110] or [-110] directions. [110] direction turns to be the easy
axis (\textit{EA}) of magnetization for all annealed samples,
while the as-grown samples show an \textit{EA} along [-110]
\cite{Stanciu:Anis}. Fig.~\ref{fig:fig1} (b) and (c) show examples
of $M(T)$ measurements recorded along the \textit{EA} for two
different annealing temperatures.
In Fig.~\ref{fig:fig1} (a), taking \Ta\ as a parameter, the
dependence of \tc\ of a 1000-\AA-thick sample on annealing time is
shown. Two \Ta-dependent regions can be identified; the first
region is characterized by a peak in \tc\ at $\sim 4$ h and $7$ h
for the annealing temperatures 240~$^\circ$C and 215~$^\circ$C,
respectively, while for \Ta\ $=190\ ^\circ$C there is a gradual
increase of \tc\ with annealing time. The second region occurs for
\ta\ longer than $\sim 10$ h and is characterized, for both
\textit{high} (but below 250--260~$^\circ$C) and \textit{low} \Ta,
by a slight decrease of \tc\ (by a few $K$), more pronounced at
higher \Ta\ and practically unnoticeable at $T_{a} = 190\ ^\circ$C
and below. One may however regard this region of the
\ta-dependence as a `saturated' region provided \Ta\ is not too
high or, if \Ta\ is higher than \tg, \ta\ does not exceed certain
limits (e.g., several tens of hours at 240 $^\circ$C). For
annealing temperatures close to \tg, the slight decay of \tc\ in
the second region is negligible as compared to the results
presented in Ref.~\cite{Potashink:APL-AnnEffGaMnAs2001}. At these
\Ta's the \mni\ out-diffusion process practically ends after few
hours of annealing and what follows is merely a re-configuration
of the interface region, as we discuss below.

The inset of Fig.~\ref{fig:fig1} (a) shows the dependence of the
manganese magnetic moment, derived at magnetic saturation, on
annealing time. When calculating the magnetic moment per Mn atom,
we do not consider the change of the Mn concentration within the
bulk of the \compB\ layer, which results from the
annealing-induced depletion process. If this effect was
considered, the magnetic moments would be close to the expected $4
\mu_{B}/$Mn-atom \cite{Korzhavyi:PRL-AsgaDef2002}. As one may
notice, there is a monotonous increase of m$_{sat}$ without any
special features at \ta~corresponding to the \tc-peaks.
\begin{figure}
        \centering
        \includegraphics[width=0.48\textwidth]{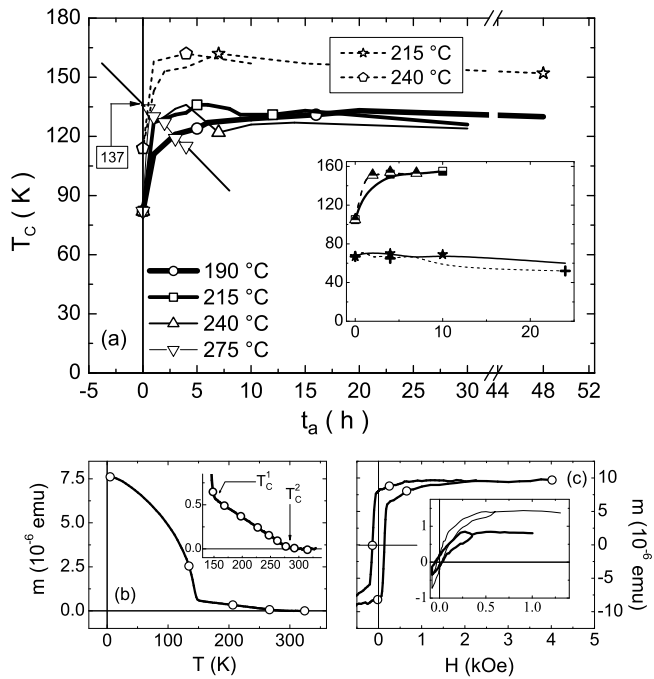}
        \caption{\label{fig:fig2} (a) \tc\ vs. annealing time for
        300 \AA (solid lines) and 500 \AA (dashed lines) thick samples
        with $x\sim6\%$ Mn. The inset shows a comparison between
        two 400 \AA~thick samples, capped with \compC\ and GaAs,
        as indicated. (b) $M(T)$ for one of the \compC~capped 400-\AA-thick samples annealed at
        275 $^\circ$C for 15 minutes. The inset shows a zoomed
        portion of the $M(T)$ curve. (c) $M(H)$ at $T=5$ K
        for the same sample as in (b), the corresponding results at $T=200$ K and 250 K
        are shown in the inset.}
\end{figure}
Due to the nonequilibrium growth conditions of LT-MBE, the
top-surface of the \compB\ layer becomes enriched in Mn, mostly
\mni, during the deposition
\cite{Yu:APL-EffThickMnI2005,Pethukov:PRL-SelfComp2002}. This is
also confirmed by our ToF-SIMS measurements on as-grown samples;
the Mn depth-profile reveals an increase in manganese
concentration close to the \compC/\compB~interface. In subsequent
annealings, the excess Mn at the interface may play the role of
nucleation centers for a presumed island growth of a Mn-As
compound. At sufficiently large \Ta, four processes may be
identified; (\textit{1}) diffusion of \mni\ from the film towards
the substrate interface (a process that does not enhance the
ferromagnetic properties as the GaAs cannot passivate \mni) [see
inset of Fig.~\ref{fig:fig2}~(a), where it is shown that
LT-annealing has no significant effect on a \compB~film capped
with 100 \AA~of GaAs], (\textit{2}) diffusion towards the film/cap
interface, (\textit{3}) migration of \mni\ at the interface, and
(\textit{4}) an eventual diffusion of \mni\ into the \compC\ cap.
In parallel, desorption of the \compC\ cap takes place, at a rate
that is also \Ta-dependent. \mni's reaching the film/cap interface
are passivated by reacting with the \compC\ cap. Initially,
heterogeneous, non-stoichiometric 2D-islands of Mn-As are formed.
During the first few hours of annealing, these islands grow
laterally rather than vertically, while for longer annealing
times, close to the \ta\ peak-values ($t_a^{peak}$), diffusion of
\mni\ into the \compC\ cap may also take place. Thus, the
\compA/\compC\ interface has a double role; (\textit{i}) it
efficiently accommodates out-diffused \mni's and (\textit{ii}) for
a certain interface coverage, it rises, though `artificially',
\tc\ of the underlying structure, hence leading to the appearance
of the \tc-peaks. However, as the \compC\ cap desorbs, the MnAs
interface compound decomposes and a manganese-oxide layer is
gradually formed. The initial `equilibrium', which brought the
\tc\ to the peak-values, is now broken and this is signaled by the
decrease of \tc\ within 1-2 hours after the $t_a^{peak}$'s. A new
steady state at the surface is reached, corresponding to the
second \ta-region mentioned above, this time in the presence of a
comparably thick manganese-oxide/arsenic-oxide layer characterized
by an almost constant value of \tc. It should be emphasized that
the drop in \tc\ is not due to O$_{2}$ or N$_{2}$ contamination,
since it has been shown in other LT-annealing studies that both
O$_{2}$ and N$_{2}$ have a positive effect on the magnetic
properties
\cite{Ku:APL-HighTcGaMnAs2003,Edmonds:MnIDiffusion,WVanRoy:cm-AnnInO2004},
but is due to the partial `alteration' of the particular state
created by the MnAs structure at the interface. A few more
arguments in favor of the above statements can be given:
(\textit{i}) The saturation magnetic moment does not change after
the first few hours of annealing, implying that the annealing
process (bulk diffusion) ends rather quickly; (\textit{ii}) by
careful annealing at high \Ta\ for a few minutes only, the MnAs
`layer' can be `caught' intact at the interface
\cite{Victor:ExplXPS}. This `layer' may be composed of either
large islands perhaps forming a percolating network or it may be a
continuous thin MnAs layer. In Fig.~\ref{fig:fig2} (b) and (c),
the $M(T)$ and $M(H)$ results are shown for a 400~\AA~thick sample
annealed at 275 $^\circ$C for 15 minutes. As we see from the
$M(T)$ dependence, the \compB~film does not appear inhomogeneous
\cite{Mathieu:PRB-InhomThin2003}, i.e., we exclude the possibility
of having a phase separation (MnAs clusters or precipitates)
within the bulk of \compB. This is also certified by the high \tc\
value ($\sim 150\ \mathrm{K}$) obtained for the \compB\ (denoted
$T_C^{1}$) and also by the shape of the $M(H)$ curve at $T = 5
\mathrm{K}$. The latter measurement was recorded after cooling the
sample to $5 \mathrm{K}$ in an applied field of 30 kOe. There is
no indication of a presumed exchange-bias shift of the hysteresis
loop, therefore the top layer is ferromagnetic which in this case
may fit a MnAs compound. The saturation magnetization extracted
from $M(H)$ for $T > T_C^{1}$, assuming a continuous layer of MnAs
at the interface, would correspond to a layer thickness of 2 nm
\cite{Tanaka:JAP-MnAsOnGaAs1994}. A second transition ($T_C^{2}$)
occurs at $\sim 280\ \mathrm{K}$ as seen in the inset of
Fig.~\ref{fig:fig2}~(b), which is in good agreement with the
ferromagnetic ordering temperature previously observed for
nanoscale MnAs dots \cite{ono:JAP-MnAs2002}.

The situation changes for thin layers. In Fig. 2 (a), \tc\ vs.
\ta\ for a series of thin samples (300~\AA, 400\AA~ and 500~\AA)
is shown. A detailed annealing study was only performed for the
300 \AA~thick series, while for the other thicknesses the aim was
to obtain \tc-values at the already established $t_{a}^{peak}$'s.
Comparing the results for the 300 \AA~and 1000
\AA~(cf.~Fig.~\ref{fig:fig1}) thick samples, it can be seen that
the \tc-peak behavior is less pronounced for the thinner sample.
The difference between the \tc-peak values and the `plateau' value
of the second region is very small. Moreover, the effect of
\textit{high} \Ta\ is stronger causing a more accelerated
degradation of the bulk \compB~ferromagnetic properties. One may
therefore conclude that thinner samples are more quickly depleted
from \mni~and that the special interface state appearing as a
result of the depletion exhibits [Mn]-dependent properties. This
implies that for a more thick sample, the number of \mni~reaching
the interface is larger, yielding larger 2D-islands of MnAs and
for thick enough samples a full layer of MnAs is formed at the
interface.

Some samples presented in Fig.~\ref{fig:fig1} were subjected to
ToF-SIMS and XPS analyses. First, the as-grown surface was
investigated. For \Ta$=215 ^\circ$C and $t_a \leq 3$ h, the XPS
results indicate only the presence of elemental As and As-O at the
top surface. However, for $t_{a} = 7$ h, Mn(2p) and Mn(3p) peaks
were identified as coming from Mn-O and Mn-As. Due to expected
small peak-shifts, it is difficult to separate the Mn-O/Mn-As
peaks, but it is clear from the $M(T)$ results (cf. e.g.
Fig.~\ref{fig:fig2} (b)) that MnAs is also present. Even after 12
h of annealing the elemental As peak is still present, though at a
smaller concentration as compared to the case of shorter \ta.
There is a weak Ga peak, indicating a desorbed cap and a thickness
of the cap remains comparable to the XPS photoelectron escape
depth, $\sim 40$ \AA. The intensity of the Mn-O/Mn-As peak does
not change much in comparison to the up-or-down change of the
intensity for the As-O and As-As peaks - the longer the \ta\ the
smaller the As-As peak becomes while the As-O peak increases in
magnitude.

\begin{figure}
        \centering
        \includegraphics[width=0.48\textwidth]{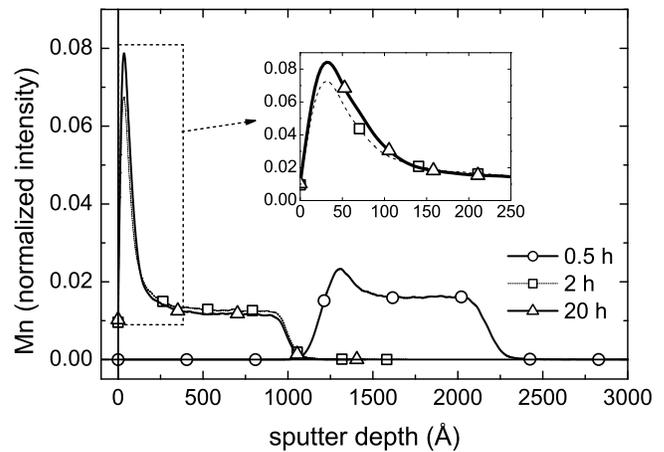}
        \caption{\label{fig:fig3} Mn depth-profiles for $1000$ \AA~thick, \compC~capped
\compB~films. The different curves correspond to different \ta\ at
$T_{a}=240 ^\circ$C; circles $t_a=0.5\ \mathrm{h}$, squares $20\
\mathrm{h}$ and, triangles $t_a=2\ \mathrm{h}$.}
\end{figure}

Manganese depth-profiles obtained from ToF-SIMS measurements on
$1000$ \AA~thick, \compC~capped \compB~films are shown in
Fig.~\ref{fig:fig3}; the different curves refer to different \ta\
at $T_{a}=240 ^\circ$C. The depth profiles clearly show that
\mni~exhibits a net diffusion towards the \compC/\compB~interface
and that \mni~is passivated at this interface. For $T_{a}=240
^\circ$C, the cap is completely desorbed after $\sim 1-2$ h of
annealing, while it takes $\sim 3-4$ h of annealing at $T_{a}=215
^\circ$C for the cap to desorb. After cap desorption, there are
only minute changes in the Mn depth-profile with increasing
annealing time; the thickness of the passivation layer, consisting
of Mn-O/Mn-As, is estimated to be $\sim 60$ \AA~from the
full-width half-maximum of Mn-peak at small sputtering depth.
The stability of the \tc-peaks is to some extent limited by the
thickness of the \compC\ cap. This can be an explanation for the
difference between the results presented here and the results of
Ref.~\cite{Adell:APL}, where slightly different growth conditions
were employed (e.g., a smaller cap thickness was used). The
apparent positive effect of the MnAs layer manifests only by a
small increase of the carrier concentration, as the Mn dopant
concentration decreases during the annealing process.

In summary, the influence of different annealing parameters on
\tc\ of \compB\ epilayers (different thicknesses but constant $x$)
has been described. For annealing temperatures close to \tg\,
relatively short annealing times are required for a complete \mni\
depletion of both thin and thick samples. This proves the
efficiency of the \compC\ cap and emphasizes the crucial role
played by the \compC/\compB\ interface in controlling both the
diffusion process within the bulk of \compB\ and the peculiar
behavior of the magnetic properties that follows after the
completion of this diffusion process. Moreover, it has been shown
that; (\emph{i}) For a particular choice of the annealing
parameters a ferromagnetic MnAs structure can be obtained at the
interface; (\emph{ii}) the impairment of the ferromagnetic
properties is much less pronounced while using the \compC-cap as
compared to case of `free' \compB~surfaces exposed either to air,
N$_{2}$ or O$_{2}$ during the LT-annealing process.

The Swedish Foundation for Strategic Research (SSF) and the Swedish
Research Council (VR) are acknowledged for financial support.
\bibliography{EnhancedTcIn_a-AsCapped_D1}

\begin{thebibliography}{28}
\expandafter\ifx\csname natexlab\endcsname\relax\def\natexlab#1{#1}\fi
\expandafter\ifx\csname bibnamefont\endcsname\relax
  \def\bibnamefont#1{#1}\fi
\expandafter\ifx\csname bibfnamefont\endcsname\relax
  \def\bibfnamefont#1{#1}\fi
\expandafter\ifx\csname citenamefont\endcsname\relax
  \def\citenamefont#1{#1}\fi
\expandafter\ifx\csname url\endcsname\relax
  \def\url#1{\texttt{#1}}\fi
\expandafter\ifx\csname urlprefix\endcsname\relax\def\urlprefix{URL }\fi
\providecommand{\bibinfo}[2]{#2}
\providecommand{\eprint}[2][]{\url{#2}}

\bibitem[{\citenamefont{Ohno and Matsukura}(2001)}]{Ohno:SolidStComRev}
\bibinfo{author}{\bibfnamefont{H.}~\bibnamefont{Ohno}} \bibnamefont{and}
  \bibinfo{author}{\bibfnamefont{F.}~\bibnamefont{Matsukura}},
  \bibinfo{journal}{Solid State Commun.} \textbf{\bibinfo{volume}{117}},
  \bibinfo{pages}{179} (\bibinfo{year}{2001}).

\bibitem[{\citenamefont{MacDonald et~al.}(2005)\citenamefont{MacDonald,
  Schiffer, and Samarth}}]{Samarth:NM-RevGaMnAs2005}
\bibinfo{author}{\bibfnamefont{A.~H.} \bibnamefont{MacDonald}},
  \bibinfo{author}{\bibfnamefont{P.}~\bibnamefont{Schiffer}}, \bibnamefont{and}
  \bibinfo{author}{\bibfnamefont{N.}~\bibnamefont{Samarth}},
  \bibinfo{journal}{Nature Materials} \textbf{\bibinfo{volume}{4}},
  \bibinfo{pages}{195} (\bibinfo{year}{2005}).

\bibitem[{\citenamefont{Liu et~al.}(1995)\citenamefont{Liu, Prasad, Weber,
  Liliental-Weber, and Walukiewicz}}]{Liu:APL-NativeDefLTGaAs1995}
\bibinfo{author}{\bibfnamefont{X.}~\bibnamefont{Liu}},
  \bibinfo{author}{\bibfnamefont{A.}~\bibnamefont{Prasad}},
  \bibinfo{author}{\bibfnamefont{E.~R.} \bibnamefont{Weber}},
  \bibinfo{author}{\bibfnamefont{Z.}~\bibnamefont{Liliental-Weber}},
  \bibnamefont{and}
  \bibinfo{author}{\bibfnamefont{W.}~\bibnamefont{Walukiewicz}},
  \bibinfo{journal}{Appl. Phys. Lett.} \textbf{\bibinfo{volume}{67}},
  \bibinfo{pages}{279} (\bibinfo{year}{1995}).

\bibitem[{\citenamefont{Korzhavyi et~al.}(2002)\citenamefont{Korzhavyi,
  Abrikosov, Smirnova, Bergqvist, Mohn, Mathieu, Svedlindh, Sadowski, Isaev,
  Vekilov et~al.}}]{Korzhavyi:PRL-AsgaDef2002}
\bibinfo{author}{\bibfnamefont{P.~A.} \bibnamefont{Korzhavyi}},
  \bibinfo{author}{\bibfnamefont{I.~A.} \bibnamefont{Abrikosov}},
  \bibinfo{author}{\bibfnamefont{E.~A.} \bibnamefont{Smirnova}},
  \bibinfo{author}{\bibfnamefont{L.}~\bibnamefont{Bergqvist}},
  \bibinfo{author}{\bibfnamefont{P.}~\bibnamefont{Mohn}},
  \bibinfo{author}{\bibfnamefont{R.}~\bibnamefont{Mathieu}},
  \bibinfo{author}{\bibfnamefont{P.}~\bibnamefont{Svedlindh}},
  \bibinfo{author}{\bibfnamefont{J.}~\bibnamefont{Sadowski}},
  \bibinfo{author}{\bibfnamefont{E.~I.} \bibnamefont{Isaev}},
  \bibinfo{author}{\bibfnamefont{Y.~K.} \bibnamefont{Vekilov}},
  \bibnamefont{et~al.}, \bibinfo{journal}{Phys. Rev. Lett.}
  \textbf{\bibinfo{volume}{88}}, \bibinfo{pages}{187202}
  (\bibinfo{year}{2002}).

\bibitem[{\citenamefont{Glas et~al.}(2004)\citenamefont{Glas, Patriarche,
  Largeau, and Lemaitre}}]{Glas:PRL-DetMnIAndAsga2004}
\bibinfo{author}{\bibfnamefont{F.}~\bibnamefont{Glas}},
  \bibinfo{author}{\bibfnamefont{G.}~\bibnamefont{Patriarche}},
  \bibinfo{author}{\bibfnamefont{L.}~\bibnamefont{Largeau}}, \bibnamefont{and}
  \bibinfo{author}{\bibfnamefont{A.}~\bibnamefont{Lemaitre}},
  \bibinfo{journal}{Phys. Rev. Lett.} \textbf{\bibinfo{volume}{93}},
  \bibinfo{pages}{086107} (\bibinfo{year}{2004}).

\bibitem[{\citenamefont{Maca and Masek}(2002)}]{Maca:PRB-MnI-MnGa2002}
\bibinfo{author}{\bibfnamefont{F.}~\bibnamefont{Maca}} \bibnamefont{and}
  \bibinfo{author}{\bibfnamefont{J.}~\bibnamefont{Masek}},
  \bibinfo{journal}{Phys. Rev. B} \textbf{\bibinfo{volume}{65}},
  \bibinfo{pages}{235209} (\bibinfo{year}{2002}).

\bibitem[{\citenamefont{Yu et~al.}(2002)\citenamefont{Yu, Walukiewicz,
  Wojtowicz, Kuryliszyn, Liu, Sasaki, and Furdyna}}]{Yu:PRB-LocOfMni2002}
\bibinfo{author}{\bibfnamefont{K.~M.} \bibnamefont{Yu}},
  \bibinfo{author}{\bibfnamefont{W.}~\bibnamefont{Walukiewicz}},
  \bibinfo{author}{\bibfnamefont{T.}~\bibnamefont{Wojtowicz}},
  \bibinfo{author}{\bibfnamefont{I.}~\bibnamefont{Kuryliszyn}},
  \bibinfo{author}{\bibfnamefont{X.}~\bibnamefont{Liu}},
  \bibinfo{author}{\bibfnamefont{Y.}~\bibnamefont{Sasaki}}, \bibnamefont{and}
  \bibinfo{author}{\bibfnamefont{J.~K.} \bibnamefont{Furdyna}},
  \bibinfo{journal}{Phys. Rev. B} \textbf{\bibinfo{volume}{65}},
  \bibinfo{pages}{201303} (\bibinfo{year}{2002}).

\bibitem[{\citenamefont{Erwin and Petukhov}(2002)}]{Pethukov:PRL-SelfComp2002}
\bibinfo{author}{\bibfnamefont{S.~C.} \bibnamefont{Erwin}} \bibnamefont{and}
  \bibinfo{author}{\bibfnamefont{A.~G.} \bibnamefont{Petukhov}},
  \bibinfo{journal}{Phys. Rev. Lett.} \textbf{\bibinfo{volume}{89}},
  \bibinfo{pages}{227201} (\bibinfo{year}{2002}).

\bibitem[{\citenamefont{Blinowski and
  Kacman}(2003)}]{Blinowski:PRB-SpinMnI2003}
\bibinfo{author}{\bibfnamefont{J.}~\bibnamefont{Blinowski}} \bibnamefont{and}
  \bibinfo{author}{\bibfnamefont{P.}~\bibnamefont{Kacman}},
  \bibinfo{journal}{Phys. Rev. B} \textbf{\bibinfo{volume}{67}},
  \bibinfo{pages}{121204} (\bibinfo{year}{2003}).

\bibitem[{\citenamefont{Wang et~al.}(2004)\citenamefont{Wang, Edmonds, Campion,
  Gallagher, Farley, Foxon, Sawicki, Boguslawski, and
  Dietl}}]{Wang:JAP-95-6512-2004}
\bibinfo{author}{\bibfnamefont{K.~Y.} \bibnamefont{Wang}},
  \bibinfo{author}{\bibfnamefont{K.~W.} \bibnamefont{Edmonds}},
  \bibinfo{author}{\bibfnamefont{R.~P.} \bibnamefont{Campion}},
  \bibinfo{author}{\bibfnamefont{B.~L.} \bibnamefont{Gallagher}},
  \bibinfo{author}{\bibfnamefont{N.~R.~S.} \bibnamefont{Farley}},
  \bibinfo{author}{\bibfnamefont{C.~T.} \bibnamefont{Foxon}},
  \bibinfo{author}{\bibfnamefont{M.}~\bibnamefont{Sawicki}},
  \bibinfo{author}{\bibfnamefont{P.}~\bibnamefont{Boguslawski}},
  \bibnamefont{and} \bibinfo{author}{\bibfnamefont{T.}~\bibnamefont{Dietl}},
  \bibinfo{journal}{J. Appl. Phys.} \textbf{\bibinfo{volume}{95}},
  \bibinfo{pages}{6512} (\bibinfo{year}{2004}).

\bibitem[{\citenamefont{Edmonds et~al.}(2002)\citenamefont{Edmonds, Wang,
  Campion, Neumann, Farley, Gallagher, and
  Foxon}}]{Edmonds:HighTcByResMonitAnn}
\bibinfo{author}{\bibfnamefont{K.~W.} \bibnamefont{Edmonds}},
  \bibinfo{author}{\bibfnamefont{K.~Y.} \bibnamefont{Wang}},
  \bibinfo{author}{\bibfnamefont{R.~P.} \bibnamefont{Campion}},
  \bibinfo{author}{\bibfnamefont{A.~C.} \bibnamefont{Neumann}},
  \bibinfo{author}{\bibfnamefont{N.~R.~S.} \bibnamefont{Farley}},
  \bibinfo{author}{\bibfnamefont{B.~L.} \bibnamefont{Gallagher}},
  \bibnamefont{and} \bibinfo{author}{\bibfnamefont{C.~T.} \bibnamefont{Foxon}},
  \bibinfo{journal}{Appl. Phys. Lett.} \textbf{\bibinfo{volume}{81}},
  \bibinfo{pages}{4991} (\bibinfo{year}{2002}).

\bibitem[{\citenamefont{Chiba et~al.}(2003)\citenamefont{Chiba, Takamura,
  Matsukura, and Ohno}}]{Chiba:EffLTAnnTri2003}
\bibinfo{author}{\bibfnamefont{D.}~\bibnamefont{Chiba}},
  \bibinfo{author}{\bibfnamefont{K.}~\bibnamefont{Takamura}},
  \bibinfo{author}{\bibfnamefont{F.}~\bibnamefont{Matsukura}},
  \bibnamefont{and} \bibinfo{author}{\bibfnamefont{H.}~\bibnamefont{Ohno}},
  \bibinfo{journal}{Appl. Phys. Lett.} \textbf{\bibinfo{volume}{82}},
  \bibinfo{pages}{3020} (\bibinfo{year}{2003}).

\bibitem[{\citenamefont{Ku et~al.}(2003)\citenamefont{Ku, Potashnik, Wang,
  Chun, Schiffer, Samarth, Seong, Mascarenhas, Johnston-Halperin, Myers
  et~al.}}]{Ku:APL-HighTcGaMnAs2003}
\bibinfo{author}{\bibfnamefont{K.~C.} \bibnamefont{Ku}},
  \bibinfo{author}{\bibfnamefont{S.~J.} \bibnamefont{Potashnik}},
  \bibinfo{author}{\bibfnamefont{R.~F.} \bibnamefont{Wang}},
  \bibinfo{author}{\bibfnamefont{S.~H.} \bibnamefont{Chun}},
  \bibinfo{author}{\bibfnamefont{P.}~\bibnamefont{Schiffer}},
  \bibinfo{author}{\bibfnamefont{N.}~\bibnamefont{Samarth}},
  \bibinfo{author}{\bibfnamefont{M.~J.} \bibnamefont{Seong}},
  \bibinfo{author}{\bibfnamefont{A.}~\bibnamefont{Mascarenhas}},
  \bibinfo{author}{\bibfnamefont{E.}~\bibnamefont{Johnston-Halperin}},
  \bibinfo{author}{\bibfnamefont{R.~C.} \bibnamefont{Myers}},
  \bibnamefont{et~al.}, \bibinfo{journal}{Appl. Phys. Lett.}
  \textbf{\bibinfo{volume}{82}}, \bibinfo{pages}{2302} (\bibinfo{year}{2003}).

\bibitem[{\citenamefont{Edmonds et~al.}(2004)\citenamefont{Edmonds,
  Boguslawski, Wang, Campion, Novikov, Farley, Gallagher, Foxon, Sawicki, Dietl
  et~al.}}]{Edmonds:MnIDiffusion}
\bibinfo{author}{\bibfnamefont{K.~W.} \bibnamefont{Edmonds}},
  \bibinfo{author}{\bibfnamefont{P.}~\bibnamefont{Boguslawski}},
  \bibinfo{author}{\bibfnamefont{K.~Y.} \bibnamefont{Wang}},
  \bibinfo{author}{\bibfnamefont{R.~P.} \bibnamefont{Campion}},
  \bibinfo{author}{\bibfnamefont{S.~N.} \bibnamefont{Novikov}},
  \bibinfo{author}{\bibfnamefont{N.~R.~S.} \bibnamefont{Farley}},
  \bibinfo{author}{\bibfnamefont{B.~L.} \bibnamefont{Gallagher}},
  \bibinfo{author}{\bibfnamefont{C.~T.} \bibnamefont{Foxon}},
  \bibinfo{author}{\bibfnamefont{M.}~\bibnamefont{Sawicki}},
  \bibinfo{author}{\bibfnamefont{T.}~\bibnamefont{Dietl}},
  \bibnamefont{et~al.}, \bibinfo{journal}{Phys. Rev. Lett.}
  \textbf{\bibinfo{volume}{92}}, \bibinfo{pages}{37201} (\bibinfo{year}{2004}).

\bibitem[{\citenamefont{Pross et~al.}(2004)\citenamefont{Pross, Bending,
  Edmonds, Campion, Foxon, and Gallagher}}]{Pross:JAP-InfLTAnnMicro2004}
\bibinfo{author}{\bibfnamefont{A.}~\bibnamefont{Pross}},
  \bibinfo{author}{\bibfnamefont{S.}~\bibnamefont{Bending}},
  \bibinfo{author}{\bibfnamefont{K.}~\bibnamefont{Edmonds}},
  \bibinfo{author}{\bibfnamefont{R.~P.} \bibnamefont{Campion}},
  \bibinfo{author}{\bibfnamefont{C.~T.} \bibnamefont{Foxon}}, \bibnamefont{and}
  \bibinfo{author}{\bibfnamefont{B.}~\bibnamefont{Gallagher}},
  \bibinfo{journal}{J. Appl. Phys.} \textbf{\bibinfo{volume}{95}},
  \bibinfo{pages}{3225} (\bibinfo{year}{2004}).

\bibitem[{\citenamefont{Hayashi et~al.}(2001)\citenamefont{Hayashi, Hashimoto,
  Katsumoto, and Iye}}]{Hayashi:APL-LTAnnGaMnAs2001}
\bibinfo{author}{\bibfnamefont{T.}~\bibnamefont{Hayashi}},
  \bibinfo{author}{\bibfnamefont{Y.}~\bibnamefont{Hashimoto}},
  \bibinfo{author}{\bibfnamefont{S.}~\bibnamefont{Katsumoto}},
  \bibnamefont{and} \bibinfo{author}{\bibfnamefont{Y.}~\bibnamefont{Iye}},
  \bibinfo{journal}{Appl. Phys. Lett.} \textbf{\bibinfo{volume}{78}},
  \bibinfo{pages}{1691} (\bibinfo{year}{2001}).

\bibitem[{\citenamefont{Potashnik et~al.}(2001)\citenamefont{Potashnik, Ku,
  Chun, Berry, Samarth, and Schiffer}}]{Potashink:APL-AnnEffGaMnAs2001}
\bibinfo{author}{\bibfnamefont{S.~J.} \bibnamefont{Potashnik}},
  \bibinfo{author}{\bibfnamefont{K.~C.} \bibnamefont{Ku}},
  \bibinfo{author}{\bibfnamefont{S.~H.} \bibnamefont{Chun}},
  \bibinfo{author}{\bibfnamefont{J.~J.} \bibnamefont{Berry}},
  \bibinfo{author}{\bibfnamefont{N.}~\bibnamefont{Samarth}}, \bibnamefont{and}
  \bibinfo{author}{\bibfnamefont{P.}~\bibnamefont{Schiffer}},
  \bibinfo{journal}{Appl. Phys. Lett.} \textbf{\bibinfo{volume}{79}},
  \bibinfo{pages}{1495} (\bibinfo{year}{2001}).

\bibitem[{\citenamefont{Malfait et~al.}(2005)\citenamefont{Malfait, Vanacken,
  Moshchalkov, Van~Roy, and Borghs}}]{WVanRoy:cm-AnnInO2004}
\bibinfo{author}{\bibfnamefont{M.}~\bibnamefont{Malfait}},
  \bibinfo{author}{\bibfnamefont{J.}~\bibnamefont{Vanacken}},
  \bibinfo{author}{\bibfnamefont{V.~V.} \bibnamefont{Moshchalkov}},
  \bibinfo{author}{\bibfnamefont{W.}~\bibnamefont{Van~Roy}}, \bibnamefont{and}
  \bibinfo{author}{\bibfnamefont{G.}~\bibnamefont{Borghs}},
  \bibinfo{journal}{Appl. Phys. Lett.} \textbf{\bibinfo{volume}{86}},
  \bibinfo{pages}{132501} (\bibinfo{year}{2005}).

\bibitem[{Vic({\natexlab{a}})}]{Victor:ExplTg}
\bibinfo{note}{$x$ might also dictate what annealing conditions one has to
  choose, however, because of its strong \tg\ dependence, specifying \tg\ is
  usually enough.}

\bibitem[{\citenamefont{Kowalczyk et~al.}(1981)\citenamefont{Kowalczyk, Miller,
  Waldrop, Newman, and Grant}}]{Kowalczyk:JVST-AsPass1981}
\bibinfo{author}{\bibfnamefont{S.~P.} \bibnamefont{Kowalczyk}},
  \bibinfo{author}{\bibfnamefont{D.~L.} \bibnamefont{Miller}},
  \bibinfo{author}{\bibfnamefont{J.~R.} \bibnamefont{Waldrop}},
  \bibinfo{author}{\bibfnamefont{P.~G.} \bibnamefont{Newman}},
  \bibnamefont{and} \bibinfo{author}{\bibfnamefont{R.~W.} \bibnamefont{Grant}},
  \bibinfo{journal}{J. Vac. Sci. Technol.} \textbf{\bibinfo{volume}{19}},
  \bibinfo{pages}{255} (\bibinfo{year}{1981}).

\bibitem[{\citenamefont{Heinlein et~al.}(1999)\citenamefont{Heinlein, Fimland,
  Grepstad, and Berge}}]{Heinlein:JVST-AsCap1999}
\bibinfo{author}{\bibfnamefont{C.}~\bibnamefont{Heinlein}},
  \bibinfo{author}{\bibfnamefont{B.}~\bibnamefont{Fimland}},
  \bibinfo{author}{\bibfnamefont{J.~K.} \bibnamefont{Grepstad}},
  \bibnamefont{and} \bibinfo{author}{\bibfnamefont{T.}~\bibnamefont{Berge}},
  \bibinfo{journal}{J. Vac. Sci. Technol. B} \textbf{\bibinfo{volume}{17}},
  \bibinfo{pages}{217} (\bibinfo{year}{1999}).

\bibitem[{\citenamefont{Adell et~al.}(2005)\citenamefont{Adell, Ilver, Kanski,
  Stanciu, Svedlindh, Sadowski, Domagala, Terki, Hernandez, and
  Charar}}]{Adell:APL}
\bibinfo{author}{\bibfnamefont{M.}~\bibnamefont{Adell}},
  \bibinfo{author}{\bibfnamefont{L.}~\bibnamefont{Ilver}},
  \bibinfo{author}{\bibfnamefont{J.}~\bibnamefont{Kanski}},
  \bibinfo{author}{\bibfnamefont{V.}~\bibnamefont{Stanciu}},
  \bibinfo{author}{\bibfnamefont{P.}~\bibnamefont{Svedlindh}},
  \bibinfo{author}{\bibfnamefont{J.}~\bibnamefont{Sadowski}},
  \bibinfo{author}{\bibfnamefont{J.~Z.} \bibnamefont{Domagala}},
  \bibinfo{author}{\bibfnamefont{F.}~\bibnamefont{Terki}},
  \bibinfo{author}{\bibfnamefont{C.}~\bibnamefont{Hernandez}},
  \bibnamefont{and} \bibinfo{author}{\bibfnamefont{S.}~\bibnamefont{Charar}},
  \bibinfo{journal}{Appl. Phys. Lett.} \textbf{\bibinfo{volume}{86}},
  \bibinfo{pages}{112501} (\bibinfo{year}{2005}).

\bibitem[{\citenamefont{Stanciu and Svedlindh}()}]{Stanciu:Anis}
\bibinfo{author}{\bibfnamefont{V.}~\bibnamefont{Stanciu}} \bibnamefont{and}
  \bibinfo{author}{\bibfnamefont{P.}~\bibnamefont{Svedlindh}},
  \bibinfo{note}{(to be published)}.

\bibitem[{\citenamefont{Yu et~al.}(2005)\citenamefont{Yu, Walukiewicz,
  Wojtowicz, Denlinger, Scarpulla, Liu, and Furdyna}}]{Yu:APL-EffThickMnI2005}
\bibinfo{author}{\bibfnamefont{K.~M.} \bibnamefont{Yu}},
  \bibinfo{author}{\bibfnamefont{W.}~\bibnamefont{Walukiewicz}},
  \bibinfo{author}{\bibfnamefont{T.}~\bibnamefont{Wojtowicz}},
  \bibinfo{author}{\bibfnamefont{J.}~\bibnamefont{Denlinger}},
  \bibinfo{author}{\bibfnamefont{M.~A.} \bibnamefont{Scarpulla}},
  \bibinfo{author}{\bibfnamefont{X.}~\bibnamefont{Liu}}, \bibnamefont{and}
  \bibinfo{author}{\bibfnamefont{J.}~\bibnamefont{Furdyna}},
  \bibinfo{journal}{Appl. Phys. Lett.} \textbf{\bibinfo{volume}{86}},
  \bibinfo{pages}{42102} (\bibinfo{year}{2005}).

\bibitem[{Vic({\natexlab{b}})}]{Victor:ExplXPS}
\bibinfo{note}{The {XPS} surface analysis shows only elemental {As} and {As-O}
  after {$t_{a}=20$~min} at {$T_{a}=275\ ^\circ\mathrm{C}$}.}

\bibitem[{\citenamefont{Mathieu et~al.}(2003)\citenamefont{Mathieu, Sorensen,
  Sadowski, Sodervall, Kanski, Svedlindh, Lindelof, Hrabovsky, and
  Vanelle}}]{Mathieu:PRB-InhomThin2003}
\bibinfo{author}{\bibfnamefont{R.}~\bibnamefont{Mathieu}},
  \bibinfo{author}{\bibfnamefont{B.~S.} \bibnamefont{Sorensen}},
  \bibinfo{author}{\bibfnamefont{J.}~\bibnamefont{Sadowski}},
  \bibinfo{author}{\bibfnamefont{U.}~\bibnamefont{Sodervall}},
  \bibinfo{author}{\bibfnamefont{J.}~\bibnamefont{Kanski}},
  \bibinfo{author}{\bibfnamefont{P.}~\bibnamefont{Svedlindh}},
  \bibinfo{author}{\bibfnamefont{P.~E.} \bibnamefont{Lindelof}},
  \bibinfo{author}{\bibfnamefont{D.}~\bibnamefont{Hrabovsky}},
  \bibnamefont{and} \bibinfo{author}{\bibfnamefont{E.}~\bibnamefont{Vanelle}},
  \bibinfo{journal}{Phys. Rev. B} \textbf{\bibinfo{volume}{68}},
  \bibinfo{pages}{184421} (\bibinfo{year}{2003}).

\bibitem[{\citenamefont{Tanaka et~al.}(1994)\citenamefont{Tanaka, Harbison,
  Park, Park, Shin, and Rothberg}}]{Tanaka:JAP-MnAsOnGaAs1994}
\bibinfo{author}{\bibfnamefont{M.}~\bibnamefont{Tanaka}},
  \bibinfo{author}{\bibfnamefont{J.~P.} \bibnamefont{Harbison}},
  \bibinfo{author}{\bibfnamefont{M.~C.} \bibnamefont{Park}},
  \bibinfo{author}{\bibfnamefont{Y.~S.} \bibnamefont{Park}},
  \bibinfo{author}{\bibfnamefont{T.}~\bibnamefont{Shin}}, \bibnamefont{and}
  \bibinfo{author}{\bibfnamefont{G.~M.} \bibnamefont{Rothberg}},
  \bibinfo{journal}{J. Appl. Phys.} \textbf{\bibinfo{volume}{76}},
  \bibinfo{pages}{6278} (\bibinfo{year}{1994}).

\bibitem[{\citenamefont{Ono et~al.}(2002)\citenamefont{Ono, Okabayashi,
  Mizuguchi, Oshima, Fujimori, and Akinaga}}]{ono:JAP-MnAs2002}
\bibinfo{author}{\bibfnamefont{K.}~\bibnamefont{Ono}},
  \bibinfo{author}{\bibfnamefont{J.}~\bibnamefont{Okabayashi}},
  \bibinfo{author}{\bibfnamefont{M.}~\bibnamefont{Mizuguchi}},
  \bibinfo{author}{\bibfnamefont{M.}~\bibnamefont{Oshima}},
  \bibinfo{author}{\bibfnamefont{A.}~\bibnamefont{Fujimori}}, \bibnamefont{and}
  \bibinfo{author}{\bibfnamefont{H.}~\bibnamefont{Akinaga}},
  \bibinfo{journal}{J. Appl. Phys.} \textbf{\bibinfo{volume}{91}},
  \bibinfo{pages}{8088} (\bibinfo{year}{2002}).

\end{thebibliography}
\end{document}